\documentclass[aps,pra,10pt,twocolumn,superscriptaddress,nofootinbib]{revtex4-2}
\usepackage{amsmath,amssymb}
\usepackage{graphicx} 
\usepackage{amsfonts}
\usepackage{tikz}
\usepackage[compat=1.1.0]{tikz-feynman}
\usepackage[english]{babel}
\usetikzlibrary{shapes, arrows.meta, positioning}
\usepackage[dvipsnames]{xcolor}
\usetikzlibrary{positioning,calc}\usepackage{bm}
\usepackage{mathtools}
\usepackage{tabularx}
\usepackage{lipsum}
\usepackage{datenumber}
\usepackage{ifthen}
\usepackage{soul}
\usepackage{etoolbox}
\usepackage[dvipsnames]{xcolor}
\usepackage{booktabs}
\newenvironment{mequation}
  {\begin{equation}\begin{aligned}}
  {\end{aligned}\end{equation}}

\newcounter{cutoffnum}
\setmydatenumber{cutoffnum}{2026}{10}{9}
\newcounter{todaynum}
\setdatetoday
\setmydatenumber{todaynum}{\thedateyear}{\thedatemonth}{\thedateday}
\makeatletter
\newcommand{\SetBaseTextColor}[1]{%
  \color{#1}\let\default@color\current@color
}
\newcommand{\ApplyCutoffColor}{%
  \ifthenelse{\value{todaynum} > \value{cutoffnum}}%
    {\SetBaseTextColor{RoyalBlue}}%
    {\SetBaseTextColor{black}}%
}
\makeatother
\AtBeginDocument{\ApplyCutoffColor}
\apptocmd\maketitle{\ApplyCutoffColor}{}{}
\definecolor{DarkBlue}{RGB}{0,0,150}
\PassOptionsToPackage{colorlinks=true,
  linkcolor=Bittersweet,  
  citecolor=RoyalBlue,  
  urlcolor=teal}{hyperref}
\usepackage{hyperref}
\usepackage[capitalise,noabbrev]{cleveref}
\begin{document}
\newcommand{\oscirc}[1]{\mathrel{\scalebox{0.75}{$\bigcirc$}_{#1}}}
\newcommand{\osstar}[1]{\mathrel{\circledast}_{#1}}

\title{Path integral quantization of the electromagnetic field in nonlinear dielectric 
materials}
\author{Arman Kashef}
\affiliation{Institute of Physics, University of Rostock, Albert-Einstein-Str. 23-24,
D-18059 Rostock, Germany}
\author{Oscar Perearnau Herrero}
\affiliation{Institute of Physics, University of Rostock, Albert-Einstein-Str. 23-24,
D-18059 Rostock, Germany}
\author{Alexander Szameit}
\affiliation{Institute of Physics, University of Rostock, Albert-Einstein-Str. 23-24,
D-18059 Rostock, Germany}
\author{Marco Ornigotti}
\affiliation{Faculty of Engineering and Natural Sciences, Tampere University, Korkeakoulunkatu 7, FI-33720 Tampere, Finland}
\author{Stefan Scheel}
\email{stefan.scheel@uni-rostock.de}
\affiliation{Institute of Physics, University of Rostock, Albert-Einstein-Str. 23-24,
D-18059 Rostock, Germany}


\begin{abstract}
We construct a quantum theory of light in nonlinear dielectric media with dispersion
and absorption. We employ a mesoscopic model for the light-matter interaction that
include a fourth-order nonlinearity in the material response. Quantization is performed
by constructing an effective action in a path-integral formalism by integrating out 
matter and bath degrees of freedom. We show how a nonlinear response function 
associated with Kerr nonlinearity is obtained through the model and, after full field
quantization, we derive the Feynman rules from this theory.
\end{abstract}
\maketitle

\section{Introduction}
The quantum theory light is a cornerstone of modern quantum technologies, as it
provides the theoretical foundations of light-matter interactions that form the
basis of applications in quantum computing, sensing and metrology. While the theory
is commonly used in free space scenarios, where the quantized electromagnetic field
is treated as in vacuo, there is a plethora of examples in which this is no longer
adequate. For example, nonclassical light propagating through optical elements such 
as phase shifters, beam splitters or optical fibers loses its quantum character
due to its interaction with the dielectric environment. 

This has led to the development of a quantum theory of light in dispersing and
absorbing media, starting with the pioneering work of 
Hopfield \cite{hopfield1958theory} in which a mesoscopic model of the electromagnetic 
field interacting with a harmonic-oscillator material, itself coupled to harmonic-oscillator 
bath modes, had been developed. Huttner and Barnett 
\cite{huttner_quantization_1992} then formulated a theory of the quantized electromagnetic
field by canonical quantization. The result was a quasi-free field theory, obtained by exact
diagonalization of a bilinear Hamiltonian, describing the linear interaction
of the quantized electromagnetic field with a dielectric medium,
whose optical properties are defined by the coupling parameters of the mesoscopic
model. While this first model was restricted to isotropic and homogeneous media, 
more elaborate models that incorporate spatial inhomogeneities and boundaries
\cite{Suttorp2004}, or magnetoelectric properties of matter \cite{soltani2008} have 
followed.

The next crucial step was the observation that any linearly responding materials, 
defined by its (measurable) optical properties, can be used to construct a quantum
theory of light in dispersing and absorbing media based on a source-quantity 
representation of the electromagnetic field using dyadic Green functions 
\cite{Matloob1995,Gruner1996,Dung1998,Scheel1998,Tip2001}. For more comprehensive
reviews and applications to dispersion forces and interactions with plasmonic 
nanoparticles, see 
Refs.~\cite{Scheel2008Acta,Buhmann2013I,Kaminer_review,OrnigottiBook}.

The quantum theory of light in absorbing media combines both aspects of quantum field
theory in terms of the relevant equal-time field commutators, and quantum statistics
in terms of the linear fluctuation-dissipation theorem (FDT). This implies that only
linearly responding materials can be consistently incorporated into the theory.
This restriction presently precludes one to treat nonlinear optical processes
on the same footing as absorption. Field quantization of nonlinear, but lossless media
had been achieved \cite{HilleryMlodinow1997}, and attempts have been made to derive
nonlinear noise polarizations from the linear theory 
\cite{ScheelWelsch2006,ScheelWelsch2006b}, while others have constructed effective 
nonlinear theories with linear absorption \cite{Lindel2021}. However, none of the
attempts have shown a consistent incorporation of the nonlinear FDT \cite{Heinz2002}.

Nonetheless, models for the quantization of light in absorbing media in the presence of 
nonlinearities based on the quantization scheme described above have been employed in 
different areas of quantum nonlinear and single-photon photonics, as a means to rigorously 
describe spectroscopy with undetected photons in the high-gain regime \cite{saravi}, or 
single-photon nonlinear dynamics of Kerr-type epsilon-near-zero (ENZ) media 
\cite{ornigotti2025}. For the latter, in particular, a rigorous, analytical model of the 
nonlinear light-matter interaction at the quantum level could represent an important step 
forward towards obtaining a deeper insight on the microscopic mechanisms contributing to 
the nonlinear response of ENZ materials \cite{ref46,ref47,ref48} and their innate 
non-perturbative character  \cite{reshef2017,ornigotti2024}, both of which, presently, can 
only be fully accounted using complex numerical codes based on complicated extensions of 
the hydrodynamic model \cite{scalora1,scalora2,scalora3}.

Motivated by all of this, in this work, we will construct a nonlinear extension of the 
mesoscopic model of the electromagnetic field interaction with absorbing matter
\cite{huttner_quantization_1992,Ornigotti2019}, in which we add quartic 
anharmonicities to the harmonic oscillators modelling the matter polarization
\cite{Schmidt1998}. In this way, we will derive firstly a nonlinear ($\chi^{(3)}$)
susceptibility out of the coupling constants of the mesoscopic model. Secondly,
we will perform a path integral quantization that has previously been done in the
linear theory \cite{bechler_quantum_1999} for the nonlinear model that will give rise
to nonlinear noise polarizations and will, eventually, lead to a quantum theory of 
light in nonlinear absorbing media that is fully consistent with the nonlinear FDT.

This article is organized as follows. In Sec.~\ref{sec:intro} we briefly recall the
mesoscopic Huttner-Barnett model \cite{huttner_quantization_1992} and the effective
action resulting from it \cite{bechler_quantum_1999}. In Sec.~\ref{sec:self} we 
construct the generating function of the nonlinear model that forms the basis for
the field quantization, followed by the Feynman rules of the theory in 
Sec.~\ref{sec:feynman}. The central quantity, the quantized displacement field, 
is derived in Sec.~\ref{sec:displacement}. We summarize our results in 
Sec.~\ref{sec:conclusions}. Detailed derivations that unnecessarily interrupt the 
flow of our arguments are relegated to the Appendix.

\section{Effective action of the electromagnetic field in media in the 
Huttner-Barnett model}
\label{sec:intro}

\begin{figure}[h!]
\centering
\includegraphics[width=0.45\textwidth]{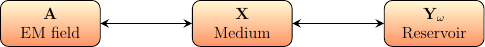}
\caption{Schematic of the Huttner-Barnett model in which the free electromagnetic
field is coupling to a harmonic oscillator field $\mathbf{X}$ modelling the 
matter degrees of freedom. They themselves are coupled to an harmonic-oscillator
reservoir $\mathbf{Y}_\omega$.}
\label{fig:hb}
\end{figure}

The Huttner–Barnett model of electromagnetism in absorbing media
\cite{huttner_quantization_1992} provides a mesoscopic quantum electrodynamical 
framework whose coarse-grained degrees of freedom yield the mesoscopic linear-response 
behavior of a dispersive and absorbing dielectric medium.
It treats the dielectric medium as a continuous field that represents the collective 
response of many microscopic dipoles. This field couples to the electromagnetic fields 
as well as to a reservoir of oscillators (see Fig.~\ref{fig:hb}). The model gives a 
full quantum description of light in linearly responding media with the material and 
reservoir degrees of freedom both modeled as harmonic oscillators. The interactions 
between the different sectors are implemented as linear bijections between their field 
sectors, rendering the theory quasi-free. Consequently, the dynamics of the 
electromagnetic field is governed by linear response only and does not include 
nonlinear effects.

In this model, the harmonic oscillator field describing the material response
singles out the dominant interaction with an external electromagnetic driving field
\cite{hopfield1958theory}.
The reservoir degrees of freedom, which may include phonons as well as other
electromagnetic field modes, account for dissipative processes in the spirit of
the open-system approach developed by Feynman and Vernon \cite{feynman_vernon_1963}.
The introduction of the reservoir provides a mechanism for dissipation by allowing 
energy to flow irreversibly out of the system (the matter oscillator), while at the 
same time ensuring that the resulting optical response function satisfies the 
Kramers–Kronig relations, which guarantee a causal electromagnetic response 
\cite{toll_1956}.

The Lagrangian density of this model can be written as
\begin{equation}
    \begin{aligned}
    \mathcal{L}_{_{\text{HB}}}[\mathbf{A},\mathbf{X},\mathbf{Y_{\omega}}]&=\mathcal{L}_{\text{e.m.}}[\mathbf{A}]+\mathcal{L}_{\text{mat}}[\mathbf{X}]+\mathcal{L}_{\text{res}}[\mathbf{Y_{\omega}}]\\ & \quad +\mathcal{L}_{\text{e.m.-mat}}[\mathbf{A},\mathbf{X}]+\mathcal{L}_{\text{mat-res}}[\mathbf{X},\mathbf{Y}_{\omega}],
    \end{aligned}
\end{equation}
where $\mathcal{L}_{\text{e.m.}}[\mathbf{A}]$ stands for the Lagrangian density of
the free electromagnetic field 
\begin{equation}
    \mathcal{L}_{\text{e.m.}}[\mathbf{A}] = \frac{\varepsilon_0}{2} \dot{\mathbf{A}}^2 - \frac{1}{2\mu_0} (\nabla \times \mathbf{A})^2,
\end{equation}
where the Weyl gauge ($\phi=0$), in which only the vector potential remains 
a dynamical variable, is chosen.  The medium polarization is modeled by the harmonic oscillator 
field $\mathbf{X}$ of frequency $\omega_0$ related to Lorentz oscillator model   
\begin{equation}
    \mathcal{L}_{\text{mat}}[\mathbf{X}] =\frac{g(\bm{x})}{2\varepsilon_0 \omega_0^2 \chi_s} 
\left( \dot{\mathbf{X}}^2 - \omega_0^2 \mathbf{X}^2 \right),
\end{equation} 
where $\omega_{0}$ is the eigenfrequency of the harmonic oscillator, $\chi_s$ is a 
dimensionless coefficient that can be interpreted as the static susceptibility
($\chi_s = \chi^{(1)}(\omega=0)$) of the medium, quantifying the stiffness 
of the bound charges at zero frequency. The characteristic function $g(\bm x)$ instead accounts
for the spatial inhomogeneity of the material. In the case of the homogeneous dielectric 
material of finite extent, for example, it defines the boundary between the dielectric and its 
surroundings, so that $g(\bm x)=1$ inside the material, and $g(\bm x)=0$ outside of it.

To model dissipation, the matter polarization field is coupled to a continuum of 
harmonic oscillators $\mathbf{Y}_\omega(\bm x,t)$ with frequency $\omega$ and mass 
density per unit frequency $\rho(x)$
\begin{equation}
    \mathcal{L}_{\text{res}}[\mathbf{Y}_\omega] = 
 \int_0^\infty d\omega' \frac{\rho}{2}\left(
 \dot{\mathbf{Y}}_{\omega'}^2 - 
\omega'^2 \mathbf{Y}_{\omega'}^2
\right).
\end{equation}
The material polarization field couples to the electromagnetic field through an 
interaction term 
\begin{equation}
    \mathcal{L}_{\text{e.m.-mat}}[\mathbf{A}, \mathbf{X}] = h(\bm{x})
 \mathbf{\dot{A}} \cdot \mathbf{X} ,
\end{equation}
where $h(\bm x)= \alpha g(\bm x)$ and $\alpha $ is the coupling constant. 

Furthermore, the interaction with the reservoir is expressed as
 \begin{equation}
     \mathcal{L}_{\text{mat-res}}[\mathbf{X}, \mathbf{Y}_\omega] = 
- g(\bm{x}) \int_0^\infty d\omega' \, \nu(\omega') \, 
\mathbf{X} \cdot \dot{\mathbf{Y}}_{\omega'},
 \end{equation}
with a frequency-dependent coupling function $\nu(\omega')$. 

The bilinear structure of the Huttner-Barnett model (as a quasi-free field theory)
lends itself to exact diagonalization \cite{huttner_quantization_1992}. Alternatively,
it can be naturally expressed in terms of an action and be formulated within the 
path-integral framework. In this framework, Bechler provided an alternative approach 
for deriving both the microscopic properties of the theory and its macroscopic 
electromagnetic response \cite{bechler_quantum_1999}. The functional integral 
formalism offers a convenient platform for this purpose, as the various field degrees 
of freedom can be integrated out successively to generate an effective action. This is 
a classical functional of the remaining fields that nevertheless incorporates the 
quantum effects of those that have been integrated out.

The effective generating function is then written as
\begin{equation}
    Z_L^{\text{eff}}[\mathbf{A}] = \int \mathcal{D}\mathbf{X}\, \mathcal{D}\mathbf{Y}_\omega \,
e^{\frac{i}{\hbar} S_{_{HB}}[\mathbf{A} ,\mathbf{X},\mathbf{Y}_\omega]}
\coloneqq e^{\frac{i}{\hbar} S_L^{\text{eff}}[\mathbf{A}]}.
\label{eq:definition_Z_eff}
\end{equation}
By integrating out the matter fields $\mathbf{X}$ and $\mathbf{Y}_\omega$, their 
effects on the electromagnetic field are absorbed into the effective action 
$S_L^{\text{eff}}[\mathbf{A}]$.  This procedure transforms the microscopic 
light-matter description into an effective macroscopic one, where the electromagnetic field 
propagates in a dressed vacuum characterized by a frequency-dependent dielectric 
response. 

The functional integration in this step is quite straightforward because both 
$\mathbf{X}$ and $\mathbf{Y}_\omega$ appear quadratically and the integrals reduce to 
Gaussian form. The Gaussian integrations result in an effective action
\begin{equation}
\begin{aligned}
    S_L^{\text{eff}}&[\mathbf{A}] = S_{\text{e.m.}}[\mathbf{A}]\\ &  + \frac{1}{2} \int dt\,dt'\,d^3x \,{\alpha^2}g(\bm{x})\,
  \dot{\mathbf{A}}(t,\bm{x})\,
  \Gamma(t-t',\bm{x})\,
  \dot{\mathbf{A}}(t',\bm{x}),
\end{aligned}
\end{equation}
where $S_\text{e.m.}$ represents the usual free electromagnetic action. The effective action results from adding an additional term in which the memory kernel (temporal Green function) 
$\Gamma(t-t',\bm x)$ is a rank-2 tensor that represents a correction to the free 
electromagnetic field propagator. It introduces temporal nonlocality (retardation) 
into the electromagnetic effective action and encodes dispersion and absorption properties of the 
dielectric medium \cite{bechler_quantum_1999}.

This constitutes the basics of the linear theory. However, numerous physical 
situations involve nonlinear effects that cannot be captured within this framework. 
In the following, we generalize the Huttner–Barnett model to include nonlinear 
interactions, thereby establishing a consistent quantum field–theoretic description 
of nonlinear optical phenomena in dissipative media. 

\section{Nonlinear extension of the Huttner-Barnett model}
\label{sec:self}

Throughout this work we adopt the formulation of the electromagnetic field in terms 
of potentials \cite{Fermi1932}. This choice is motivated by the fact that the gauge 
freedom is manifest within this representation. This puts us in a position where we 
can find a suitable way to fix the gauge and thus remove the unphysical degrees of 
freedom. A sensible choice is the Weyl gauge which eliminates the scalar potential in 
the eventual functional integration over the field configurations 
\cite{Weyl1929,ORaifeartaigh2000}.
By contrast, a formulation purely in terms of $\mathbf{E}$ and $\mathbf{B}$ would 
involve integrating over two vector fields subject to non-trivial
constraints that introduce nonlocal couplings between different field values, thus 
breaking locality. The formulation in terms of potentials solves this and leads to a 
systematically gauge-fixed description of the quantum field.

The aim is to extend the previously discussed model to include nonlinear effects
\cite{Bloembergen1965}. Macroscopically, the electromagnetic field propagates in the 
medium through an anharmonic effective potential. Within the Huttner–Barnett 
framework, that potential is created by the matter field. Consequently, we model the 
nonlinear response as a self-interaction of the matter field, which we thus regard as
the microscopic origin of the macroscopic nonlinearity.
The simplest extension that preserves inversion symmetry is to modify the model such that the 
field $\mathbf{X}$ 
exhibits a quartic contribution of the form $\lambda \mathbf{X^{4}}$ 
\cite{ItzyksonZuber1980}. This corresponds, at the level of equations of motion, to the Duffing 
oscillator model, frequently used in nonlinear optics to model third-order nonlinear 
interactions, i.e., third-order susceptibilities \cite{boyd_nonlinear_2020}. 
As the fields are real in the time domain, their Fourier transforms are complex fields in
the frequency domain. The common choice in interacting field theories with complex fields is 
\begin{equation}
\begin{aligned}
S_{\lambda}[\mathbf{X}]
&= \int \! d^{3}x\,
   \int\limits_0^\infty \frac{d\omega_{1}d\omega_{2}d\omega_{3}d\omega_{4}}{(2\pi)^{3}}
   \delta\!\bigl(\omega_{1}-\omega_{2}+\omega_{3}-\omega_{4}\bigr)
\\
&\cdot\,
   \lambda_{\mu\nu\alpha\beta}(\bm{x})\,
   X_{\mu}(\omega_1)X^{\ast}_{\nu}(\omega_2)X_{\alpha}(\omega_3)X^{\ast}_{\beta}(\omega_4).
   \label{S_lambda_freq}
\end{aligned}
\end{equation}
which is equivalent to a four-wave mixing process with two incoming and two outgoing fields. 
Note that equivalent combinations result from the symmetry of the coupling tensor 
$\lambda_{\mu\nu\alpha\beta}(\bm{x})$ with respect to pairwise exchange of indices.

The action is therefore extended accordingly to
\begin{equation}
    S_{\text{NL}}[\dot{\mathbf{A}},\mathbf{X},\mathbf{Y}_{\omega}]=S_{\text{HB}}[\dot{\mathbf{A}},\mathbf{X},\mathbf{Y}_{\omega}]+S_{\lambda}[\mathbf{X}],
\end{equation}
and the corresponding generating function, including the auxiliary source fields, of 
the nonlinear theory then reads
\begin{equation}
\begin{aligned}
Z_{\text{NL}}[\mathbf{J},\mathbf{f}] 
&= \int \mathcal{D}\mathbf{Y}_{\omega}\,\mathcal{D}\mathbf{X}\,\mathcal{D}\mathbf{A}\;
\exp\!\left\{
\frac{i}{\hbar}\Big[
S_{HB}[\mathbf{A},\mathbf{X},\mathbf{Y}_{\omega}]
\right.\\[4pt]
&\left.\qquad
+ S_{S_X}[\mathbf{X},\mathbf{f}]
+ S_{S_A}[\mathbf{A},\mathbf{J}]
+ S_{\lambda}[\mathbf{X}]
\Big]
\right\},
\end{aligned}
\end{equation}

where 
\begin{align}
    S_{S_{A}}[\mathbf{A},\mathbf{J}]&=\int \,\mathrm{d}^{3}x\,\mathrm{d}t\,\mathbf{J}\!\cdot\!\mathbf{A},\\
    S_{S_{X}}[\mathbf{X},\mathbf{f}]&=\int \,\mathrm{d}^{3}x\,\mathrm{d}t \,
g(\mathbf{x})\, \mathbf{f}\!\cdot\!\mathbf{X}.
\end{align}
In the remainder of the section, we shall systematically quantize each field through 
the use of functional integrals.

\subsection{Integration over the reservoir field}

As the coupling between the system and the reservoir remains linear, the integration 
over the reservoir variables is still Gaussian, and its contribution is identical to 
that of the linear case \cite{bechler_quantum_1999}. We denote the result of the 
integration by $\mathcal{I}_{\text{Y}}$ with
\begin{equation}
    \mathcal{I}_{Y}=\mathcal{N}_{Y}e^{\frac{i}{\hbar}S_{Y}^{\mathrm{infl}}[X]},
\end{equation}
where
\begin{mequation}
S_{Y}^{\mathrm{infl}}[X] 
&= 
    -\frac{1}{2}\,\int  d^{3}x \, dt\,
    \int_{0}^{\infty} d\omega' \frac{|\nu(\omega')|^2 g(x)}{\rho} \,\Big[
  \\
&
  \quad\mathbf{X}(t)^{2} - \int  dt'\,
    \mathbf{X}(t)\, {\omega'}^2D_F(t - t', \omega')\, \mathbf{X}(t')
\Big]
\label{influence_action}
\end{mequation}
and $\mathcal{N}_{Y}$ is a normalization constant.

Here $D_F(t-t',\omega')$ denotes the Feynman propagator (Green’s function) of the harmonic reservoir oscillators.

It is worth noting that Eq.~\eqref{influence_action} is a 
influence functional for a Gaussian bath. The 
influence action plays the role of a silent intermediary: it carries, in a compact 
form, the entire effect of the environmental degrees of freedom on the system's 
dynamics once they have been integrated out. Its local contribution renormalizes the 
bare natural frequency of the matter field, reflecting how the quantum fluctuations of 
the reservoir reshape the dynamics of the matter. The temporal non-local term, endowed 
with a memory kernel, gives rise to dissipation, imprinting the irreversibility of the 
energy flow from the system to its surroundings. Both contributions identified above 
arise from the full kernel of the reservoir, $\Sigma$, defined as
\begin{equation}
   \begin{aligned}
        \Sigma_{\mu\nu}(t-t',\bm{x})=\int_{0}^{\infty}d\omega'\, \frac{1}{\rho}\big[|\nu(\omega')|^{2} \delta(t-t')\delta_{\mu\nu}\\-\omega'^2(D_F)_{_{\mu\nu}}(t-t',\omega')\big].
   \end{aligned}
\end{equation}
This tensor modifies the free propagator of the quantized matter field. 
Then, the action can be written in the compact form
\begin{equation}
    S_{Y}^\text{infl}[\mathbf{X}]=-\frac{1}{2}\int dtdt'd^{3}x\,
    g(\bm{x})\,X_{\mu}\Sigma_{\mu\nu}X_{\nu},
\end{equation}
so that now the effective nonlinear action is
\begin{equation}
    \begin{aligned}
        &S_{NL}^{\text{eff}}[\mathbf{A},\mathbf{X},\mathbf{f},\mathbf{J}] = S_{e.m.}[\mathbf{A}]+S_{S_{A}}[\mathbf{A},\mathbf{J}]+S_{mat}[\mathbf{X}]\\ & +S_{e.m.-mat}[\mathbf{A},\mathbf{X}]\quad+S_{Y}^\text{infl}[\mathbf{X}]+S_{S_{X}}[\mathbf{X},\mathbf{f}]+S_{\lambda}[\mathbf{X}]\\&
        = S_{e.m.}[\mathbf{A}]+S_{S_{A}}[\mathbf{A},\mathbf{J}]+ S_{\text{Gauss}}[\mathbf{A}, \mathbf{X},\mathbf{f}] + S_{\lambda}[\mathbf{X}].
    \end{aligned}
\end{equation}
All the quadratic contributions of the matter field have been included into 
$S_{\text{Gauss}}$ which reads
\begin{widetext}
    \begin{equation}
 S_{\text{Gauss}}[\mathbf{A}, \mathbf{X},\mathbf{f}] 
= \int \frac{d\omega}{2\pi} \int d^3x \, g(x)\, 
 \times
    \bigg [ \frac{1}{2} X_{\mu}^{*}(\omega, \bm{x}) 
    \Bigl( \frac{\omega^{2}-\omega_{0}^{2}}{\varepsilon_0 \omega_0^2 \chi_s(x)} \delta_{\mu\nu}  - \Sigma_{\mu\nu}(\omega, \bm{x}) \Bigl)
    X_{\nu}(\omega, \bm{x}) 
 + \Bigl( f_{\mu}(\omega, \bm{x}) - {\alpha}E_{\mu}(\omega, \bm{x}) \Bigr)
     X_{\mu}(\omega, \bm{x})\bigg ].
\end{equation}

\subsection{Integration over the matter field}

Because the nonlocal term contained in the reservoir kernel corresponds to a 
convolution in the time domain, it is convenient to work in the frequency 
representation, where these convolutions become simple products.
In the sector of the generating function governed by the matter fields, two contributions arise:
a quadratic one from $S_\text{Gauss}$ and a quartic one from the interaction term. The full 
generating function now reads
\begin{equation}
    Z_{\text{NL}}[\mathbf{f},\mathbf{J}]
= \int \mathcal{D}\mathbf{A}\,
   \exp\!\left\{\frac{i}{\hbar}\!\left(
      S_{\text{e.m.}}[\mathbf{A}]
      + S_{S_{A}}[\mathbf{A},\mathbf{J}]
   \right)\right\}
   \times\int \mathcal{D}\mathbf{X}\,
   \exp\!\left\{\frac{i}{\hbar}\!\left(
      S_{\lambda}[\mathbf{X}]
      + S_{\text{Gauss}}[\mathbf{A},\mathbf{X}]
   \right)\right\}.
\end{equation}
At this stage, the integral over the matter field cannot yet be performed, as it is rendered 
non-Gaussian by the presence of $S_\lambda$.  Perturbation theory comes to the rescue, by 
replacing the quartic term by functional derivatives with respect to the source field 
$\mathbf{f}$, allowing one to extract it from the integration over $\mathbf{X}$. Thus, under 
the assumption that $\lambda \ll 1$, the generating function becomes

\begin{equation}
Z_{\text{NL}}[\mathbf{f},\mathbf{J}]= \int \mathcal{D}\mathbf{A}\,
   \exp\!\left\{\frac{i}{\hbar}\!\left(
      S_{\text{e.m.}}[\mathbf{A}]
      + S_{S_{A}}[\mathbf{A},\mathbf{J}]
   \right)\right\}  \times Z_\lambda[\mathbf{A,\mathbf{f]}}
   \int \mathcal{D}\mathbf{X}\exp\left\{\frac{i}{\hbar}S_{\text{Gauss}}[\mathbf{A},
   \mathbf{X},\mathbf{f}]\right\},
\label{eq:Z_with_derivatives}
\end{equation}

where 

\begin{mequation}
    Z_\lambda[\mathbf{A,\mathbf{f]}}= \Bigg[
      1
      + 2\pi\frac{i}{\hbar}
       & \int d^{3}x\,
\prod_{i=1}^4 
\frac{d\omega_i}{2\pi}\,
        \delta(\omega_{1}-\omega_{2}+\omega_{3}-\omega_{4})\, 
        \frac{1}{4!}
        \left(\frac{\hbar}{i}\right)^{4}
        \lambda_{\mu\nu\alpha\beta}(\bm{x})
      \frac{\delta}{\delta f_{\beta}^{*}}\,
      \frac{\delta}{\delta f_{\alpha}}\,
      \frac{\delta}{\delta f_{\nu}^{*}}\,
      \frac{\delta}{\delta f_{\mu}}
      + \mathcal{O}(\lambda^{2})
   \Bigg]\,
\end{mequation}

The integral over the matter field can now be carried out analytically. The resulting 
expression, called $Z_{\text{Gauss}}$, is
\begin{equation}
\begin{aligned}
& Z_{\text{Gauss}}[\textbf{A},\textbf{f}]=\\
& \exp\left[\frac{i}{2\hbar}\int \frac{d\omega}{2\pi}d^3x\, g(\bm{x})(\Psi_\mu^\dagger(\omega)\, 
\Gamma_{\mu \nu}(\omega)(\mathbf{I}-\sigma_x)_{\nu\rho}\, \Psi_\rho(\omega))\right]
\end{aligned}
\end{equation}
\end{widetext}
where 
$$
\Psi_\alpha(\omega) =
\begin{pmatrix}
f_\alpha(\omega,\bm{x})\\
\alpha E_\alpha(\omega,\bm{x})
\end{pmatrix},
$$ 
and $\sigma_x$ denotes the Pauli-$x$ matrix. Setting the source field $f_\mu=0$
leads to the known result \cite{bechler_quantum_1999} after the quantization of the matter field.\\
The tensor $\Gamma$ has the full dynamical response of the composite system formed by matter and its reservoir. It captures the quantum nature of both sectors and summarizes their influence in a single frequency dependent object. Its explicit form is:
\begin{equation}
    \Gamma(\omega)=\frac{\varepsilon_0 \omega_0^2 \chi_s}{
\omega_0^2 - \omega^2 - \omega^2 \omega_0^2 \varepsilon_0\chi_s \Sigma(\omega)},
\end{equation}
where  $\Sigma(\omega)$ is the reservoir kernel in the Fourier space.\\
It is now possible to perform the functional derivatives over $Z_\text{Gauss}$. Upon using 
renormalization arguments and only keeping the connected contributions (see 
Appendix~\ref{sec:Functional_derivatives}), the result can be cast as a polynomial functional 
leading to
\begin{mequation}
Z_{\mathrm{NL}}[\mathbf{f},\mathbf{J}]
= \int \mathcal{D}\mathbf{A}\,
  & \exp\!\left\{\frac{i}{\hbar}\!\left(
      S_{\text{e.m.}}[\mathbf{A}] + S_{S_{A}}[\mathbf{A},\mathbf{J}]
   \right)\right\}\\  
   &\times \left[1+\int \tilde{d}\omega P[\mathbf{A,f}]\right]\,
   Z_{\text{Gauss}}[\mathbf{A},\mathbf{f}],
\label{eq:Zeff}
\end{mequation}
with the definition
\begin{equation}
    \begin{aligned}
       P[\mathbf{A},\mathbf{f}]&= 
       \Lambda_{\alpha\beta\mu\nu}(\bm{x})\,E_{\alpha}\,E^{*}_{\beta}\,E_{\nu}\,E^{*}_{\mu}\\
        &- 2\big(\Delta_{\beta\nu\mu}(\bm{x})\,E^{*}_{\beta}\,E_{\nu}\,E^{*}_{\mu}+\text{c.c.}\big)\\[4pt]
&+ 4\,\Phi_{\nu\mu}^{(1)}(\bm{x})\,E_{\nu}\,E^{*}_{\mu}
        + \big(\Phi_{\nu\mu}^{(2)}(\bm{x})\,E^{*}_{\nu}\,E^{*}_{\mu}+\text{c.c.}\big)\\
        &- 2\big(\Xi_{\mu}(\bm{x})\,E_{\mu}^{*}+\text{c.c.}\big)+ \Lambda^{(0)}_{\alpha\beta\mu\nu}(\bm{x})\,f_{\alpha}\,f_{\beta}^{*}\,f_{\nu}\,f_{\mu}^{*}
        ,
    \end{aligned}
\label{Z_poly}
\end{equation}
and the abbreviation
$$\int \tilde{d}\omega \coloneqq \frac{i}{16\hbar}\idotsint d^3x \displaystyle\prod_{i=1}^4 
\frac{d\omega_i}{2\pi}\delta(\omega_1-\omega_2+\omega_3-\omega_4).$$
Moreover, the tensors in Eq.~(\ref{Z_poly}) are abbreviations of the following compound 
expressions: 
\begin{flalign}
\!\!\!\!\!\!\!
\Lambda_{\alpha\beta\mu\nu}^{(0)}(\bm{x})
&= \frac{g^{4}(\bm{x})}{4!}
\lambda_{\rho\sigma\xi\gamma}(\bm{x})
\Gamma_{\alpha\rho}(\bm{x})
\Gamma_{\beta\sigma}(\bm{x})
\Gamma_{\mu\xi}(\bm{x})
\Gamma_{\nu\gamma}(\bm{x}),\\ \label{eq:Lambda}
\Lambda_{\alpha\beta\mu\nu} &= \Lambda_{\alpha\beta\mu\nu}^{(0)} \alpha^4,\\
\Delta_{\beta\mu\nu}(\bm{x}) &= \Lambda_{\alpha\beta\mu\nu}^{(0)}\alpha^3 f_{\alpha},\\
\Phi^{(1)}_{\mu\nu}(\bm{x}) &= \Lambda_{\alpha\beta\mu\nu}^{(0)} \alpha^2 
f_{\alpha}f^{*}_{\beta},\\
\Phi^{(2)}_{\mu\nu}(\bm{x}) &= \Lambda_{\alpha\beta\mu\nu}^{(0)} \alpha f_{\alpha}f_{\beta},\\
\Xi_{\nu} (\bm{x}) &= \Lambda_{\alpha\beta\mu\nu}^{(0)}f_{\alpha}f_{\beta}^{*}f_{\mu}.
\end{flalign}

Equation~(\ref{Z_poly}) places us into a position to able to compute the nonlinear susceptibility
as a function of the linear susceptibility, which we will explicitly show later.
With the self-interaction of the matter field quantized, the internal dynamics of this sector 
are now fully described within the quantum framework. We now proceed to finalize the full 
quantization of the model by applying the same procedure to the remaining classical field, the 
electromagnetic field. 

\subsection{Full quantization of the model}
Up to this point, our formulation of the electromagnetic field coupled to the medium has been 
semi-classical. By using Eq.~(\ref{eq:Zeff}) we can now start to explicitly calculate the 
functional integral for which we find 
\begin{equation}
Z_{\mathrm{NL}}[\mathbf{f},\mathbf{J}] = C_{f}[\mathbf{f}] P\left[\mathbf{f},
\frac{\delta}{\delta\mathbf{J}}\right] Z_0[\mathbf{f},\mathbf{J}],
\end{equation}
where $C_{f}[\mathbf{f}]$ collects the terms independent of $A_\mu$ and represent the self-
energy of the medium in the absence of the electromagnetic field
\begin{equation}
C_{f}[\mathbf{f}] =
\exp\!\left\{
\frac{i}{2\hbar}\,\!
\int d^3x\,\frac{d\omega}{2\pi}\,
g(\bm{x})\,
\mathbf{f}^*(\omega,\bm{x})\,
\Gamma(\omega,\bm{x})\,
\mathbf{f}(\omega,\bm{x})
\right\},
\label{eq:C_f}
\end{equation}
which, as we have seen before, contributes to $Z_\text{Gauss}$.

The functional $P$ represents the polynomial part of the functional integral Eq.~\eqref{Z_poly},
and all terms that are of higher than quadratic order in the electromagnetic field are contained 
in $P$. While the Gaussian portion of the generating function arises from the quadratic dependence 
of the action on the field and can be integrated exactly, the polynomial part encodes the 
residual nonlinear interactions or perturbative corrections to the electromagnetic dynamics.
In Eq.~(\ref{eq:C_f}), we replaced the field $\mathbf{A}$ by functional derivatives with respect 
to the source field $\mathbf{J}$ because differentiating the generating functional with respect 
to a source inserts that field into the path integral. Thus, the effect of a field can be 
reproduced by $-i\hbar \delta/\delta J$.

The functional $Z_0[\mathbf{f},\mathbf{J}]$ encompasses all the remaining terms in 
Eq.~\eqref{eq:Zeff}, which include quadratic, linear, and mixed derivative contributions 
involving the vector potential $\mathbf{A}(\omega,\bm x)$, the source field 
$\mathbf{f}(\omega,\bm x)$ and the memory kernel $\Gamma(\omega)$.\\

Although the free-field part is manifestly quadratic, we need to restore the Gaussian structure
for the mixed term $i\omega\, \mathbf{f} \,\Gamma\, \mathbf{A}$. After rearrangement, all remaining 
occurrences of the electromagnetic field appear in either quadratic or linear form and, solving 
the derived Gaussian integral, the result can be expressed schematically as  

\begin{mequation}
Z_0[\mathbf{f},\mathbf{J}]
= & \mathcal{N}\,
\exp\!\{
\frac{i}{{2}\hbar}
\int  d^3 x\, d^3 x' d\omega \\ & \times \textbf{L}^{\text{tot}\,*}(\omega,\bm x)\,
D(\omega;\bm x,\bm x')\,\textbf{L}^{\text{tot}}(\omega,\bm x') \},
\end{mequation}

where the total linear source is 
\begin{equation}
    L_{\mu}^{\text{tot}}(\omega,\bm x)=
i\omega \big[\,J_\mu(\omega,\bm x)
+\frac{\alpha}{2}\,g(\bm x)\,
\Gamma_{\mu\nu}(\omega,\bm x)\,f_{\nu}(\omega,\bm x) \big ],
\end{equation}
\begin{widetext}
and $D(\omega;\bm x,\bm x')$ acts as the Green function for the kernel $K^{\text{tot}}$, where
\begin{equation}
\begin{aligned}
    K_{\mu\nu}^{\text{tot}}(\omega) &=\varepsilon_0\,\omega^2 \delta_{\mu\nu}\,
+\frac{1}{\mu_0}(\nabla\times \nabla \times)_{\mu\nu}-g(\bm x)\,\omega^2\alpha^{2}\,\Gamma_{\mu \nu}(\omega,\bm x).
\end{aligned}
\end{equation}

Differentiating $Z_0$ with respect to the source $\mathbf{J}$ pulls down factors of the
electromagnetic field $\mathbf{A}$, and the result is the classical mean field produced by the 
source where this field insertion defined as 
\begin{align}
    m_\mu(\omega,x)&\coloneqq i\omega \int  d^3 x \, {L_\nu^{\text{tot}}}^*(\omega,\bm x) D_{\mu \nu}(\omega,\bm x),\\
      m'_\mu(\omega,x)&\coloneqq i\omega \int  d^3 x \, D_{\mu \nu}(\omega,\bm x)\,{L_\nu^{\text{tot}}}(\omega,\bm x). 
\end{align}
If we differentiate twice, two fields may appear independently or may contract to form a 
propagator and so on. This is exactly the functional form of Wick's theorem.

After acting with the functional $P$ on $Z_0$ we obtain
\begin{equation}
\begin{aligned}
    Z_{NL}[\mathbf{f},\mathbf{J}]&=C_{f}[\mathbf{f}]Z_0[\mathbf{f},\mathbf{J}] \,
    \bigg[1+ \int \frac{i}{16\hbar}\tilde{d\omega}\,\sum_{i=0}^4 P_i \bigg],
\end{aligned}
\end{equation}
where

\begin{align}
P_0 &\coloneqq \Lambda^{(0)}\mathbf{f}\,\mathbf{f}^* \mathbf{f}\, \mathbf{f}^*, \\
    P_1 &\coloneqq  -2 [\Xi \,\textbf{m}'(\omega_1)+ c.c.],\\
 P_2 &\coloneqq  4\Phi^{(1)}[\textbf{m}'(\omega_2)\textbf{m}(\omega_1)+
    \omega_1\omega_2\,D(\omega_2)\delta(\omega_2-\omega_1)] -2[\Phi^{(2)}
    \textbf{m}'(\omega_1)\textbf{m}'(\omega_2)+c.c.],\\
    P_3 &\coloneqq -\Delta[\omega_3\omega_2\,\textbf{m}'(\omega_1)D(\omega_3)\delta(\omega_3-\omega_2) +\omega_1\omega_2\,D(\omega_1)\textbf{m}'(\omega_3)\delta(\omega_1-  \omega_2)  
    +\textbf{m}'(\omega_1)\textbf{m}(\omega_2)\textbf{m}'(\omega_3)]+c.c.,\\
    P_4 &\coloneqq  
    \Lambda\big[\omega_1\omega_2\omega_3\omega_4\,D(\omega_4)D(\omega_2)\delta(\omega_2-
    \omega_1)\delta(\omega_4-\omega_3)+
    \omega_1\omega_2\omega_3\omega_4\,D(\omega_2)D(\omega_4)\delta(\omega_4-
    \omega_1)\delta(\omega_2-\omega_3) \nonumber\\ & \quad +
    \omega_4\omega_1\,D(\omega_4)\textbf{m}'(\omega_1)\textbf{m}(\omega_2)\delta(\omega_4-\omega_1)+
    \omega_2\omega_1\,D(\omega_2)\textbf{m}(\omega_3)\textbf{m}'(\omega_4)\delta(\omega_2-\omega_1) \nonumber\\ 
    &\quad +\omega_2\omega_3\,\textbf{m}'(\omega_1)D(\omega_2)\textbf{m}(\omega_3)\delta(\omega_2-\omega_3)+
    \omega_4\omega_3\,\textbf{m}'(\omega_1)\textbf{m}(\omega_2)D(\omega_4)\delta(\omega_4-
    \omega_3)+\textbf{m}'(\omega_1)\textbf{m}(\omega_2)\textbf{m}'(\omega_3)\textbf{m}(\omega_4)\big].
\end{align}
\end{widetext}
All these terms can be then used to define the polynomial functional as
\begin{equation}
    P[\mathbf{J},\mathbf{f}]=\int \tilde{d\omega}\sum_{i=0}^{4}P_{i}.
\end{equation}
It is worth pausing to remark a subtle point. Although we assume to describe an 
inversion-symmetric crystal, which would forbid certain nonlinear (e.g., cubic) terms, such 
terms do appear in the intermediate steps of the calculation when the sources are included.
Their presence can be explained by understanding that the sources shift the center of the 
Gaussian distribution for the electromagnetic field, giving a non-zero expectation value. 
Formally, this resembles the spontaneous symmetry breaking that gives rise to the Higgs 
mechanism, where the vacuum is displaced from the symmetric point acquiring a non-zero value. 
Here, however, the symmetry is not broken by a fundamental field as in the Higgs case but by 
the auxiliary sources we have manually introduced. In our context, quantization itself does not 
break the symmetry; the apparent asymmetry arises solely from the external sources. The physical 
theory is recovered only in the limit where these sources are set to zero, restoring the 
symmetry of the material. 
\section{Feynman rules}
\label{sec:feynman}

With the full generating function $Z_{NL}[\mathbf{f},\mathbf{J}]$ at hand, we are now
in a position to find Feynman rules and construct the associated Feynman diagrams. 

\subsection{Tree level diagrams}
It is convenient to restrict ourselves to the study of the theory in the frequency domain. 
The choice follows naturally due to the time-translational invariance of the quadratic 
kernel. As the propagator is its inverse, time-translational symmetry ensures that in 
the frequency space the propagator is diagonal, and different frequency components are 
decoupled. 

The quadratic structure of the theory already determines the type of the propagators that 
appear at tree level. The free parts of the electromagnetic and matter fields contribute to 
terms of the forms $\mathbf{A}\mathbf{A^*}$ and $\mathbf{X}\mathbf{X^*}$, respectively. 
Bilinear couplings between such fields also belong to the quadratic part of the action 
giving rise to mixed propagators connecting both sectors, $\mathbf{AX^*}$ and 
$\mathbf{XA^*}$. Note that all other combinations are forbidden by virtue of the Wick 
theorem. The nonlinear terms, on the other hand, have no influence on the propagators at 
this order; they simply give rise to interaction vertices, whose form is governed by the 
kernel of the nonlinear part of the theory.

We consider the connected generating functional
\begin{equation}
W = W_0+W_f - i\hbar \ln\!\left( 1 + \frac{i}{16\hbar}\,  P[\mathbf{J},\mathbf{f}]  \right),
\end{equation}
with $ W_0 = -\,i\hbar \ln Z_0[\mathbf{f},\mathbf{J}]$ and $W_f= -\,i\hbar \ln C_{f}[\mathbf{f}]$.


\begin{figure*}[t]
\centering
\includegraphics[width=0.9\textwidth]{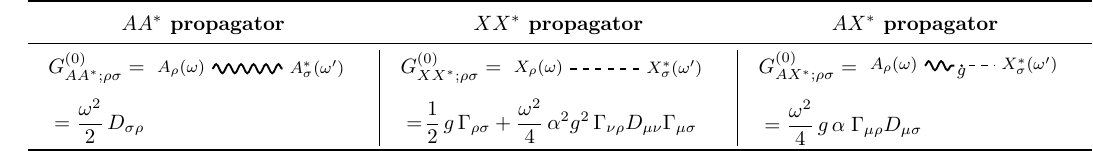}
\caption{The three propagators appearing in the theory. The superscript \((0)\) denotes the tree-level contribution.}
\label{fig:propagators}
\end{figure*}

\begin{figure*}[t]
\centering
\includegraphics[width=0.9\textwidth]{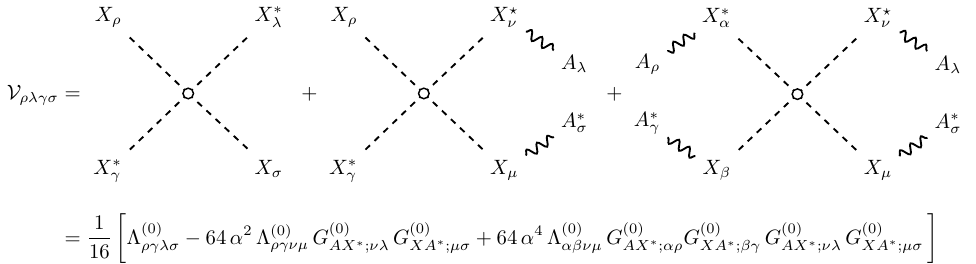}
\caption{Four-point vertex \(\mathcal{V}_{\rho\lambda\gamma\sigma}\) written as the sum of three contributions.}
\label{fig:vertex-V}
\end{figure*}
In addition, any contribution coming from the polynomial functional $P[\mathbf{J},\mathbf{f}]$
takes the form of a vertex insertion together with the corresponding loop corrections to the free 
propagator.
We can now evaluate all the Green’s functions of the theory at tree level, namely those coming 
from the free part, and represent them pictorially through their associated Feynman diagrams. The 
results are summarized in Fig.~\ref{fig:propagators}.
Similarly, the vertex can be obtained by applying four functional derivatives. In 
contrast to the propagators, only the polynomial part will affect the result and contains three 
different contributions as shown in Fig.~\ref{fig:vertex-V}.
Because the electromagnetic field and the matter field interact linearly too, the vertex unfolds 
into a superposition of processes where matter excitations may be partially replaced by 
electromagnetic ones. This multiplicity is not a decorative detail: it is precisely this linear 
intertwining that seeds the nonlinear structure of the classical theory, where the fields no 
longer propagate as isolated entities but reshape each other’s dynamics.
In this sense, the quantum vertex already contains, in compact form, the richer nonlinear behavior
that emerges at the classical level.

\subsection{Loop corrections}
As previously noted, the quartic interaction term does not contribute to the tree level 
propagators, it merely introduces a vertex term. However, due to the quantum nature of the 
theory, fluctuations are inevitably embedded in it. They correct the propagators through 
the appearance of loops. Mathematically, they appear inside the self-energy tensor. In this 
way, they ``dress" the propagator revealing how self-interactions or interactions with other 
fields reshape the propagation of particles.

The self-energy tensor is obtained by integrating the product of vertex factors and the 
internal propagator over all internal loop momenta that composes the one-particle 
irreducible ($1PI$) two-point diagram, weighted by the appropriate symmetry factor to avoid 
overcounting identical diagrams. 
In our case, there is only one ($1PI$) diagram corresponding to the bosonic vertex of 
the matter with a symmetry factor $S=1$. The bare self interaction tensor then reads as
\begin{equation}
    \begin{aligned}
        &\Pi_{\alpha\gamma}^{(\mathrm{bare})}(\omega)
        =\int\!\frac{d\Omega}{2\pi}\,
        \mathcal{V}_{\alpha\beta\mu\gamma}
        G^{(0)}_{XX^*;\beta\mu}.
\end{aligned}
\end{equation}
The linear coupling belongs to the free sector, thereby contributing only to the tree 
level propagator, so only the matter field contributes to this correction. Yet, through 
the $\mathbf{A}$-$\mathbf{X}$ interaction the self-energy effect will be present in the 
propagators between all fields.

\begin{figure}[t]
\centering
\includegraphics[width=0.45\textwidth]{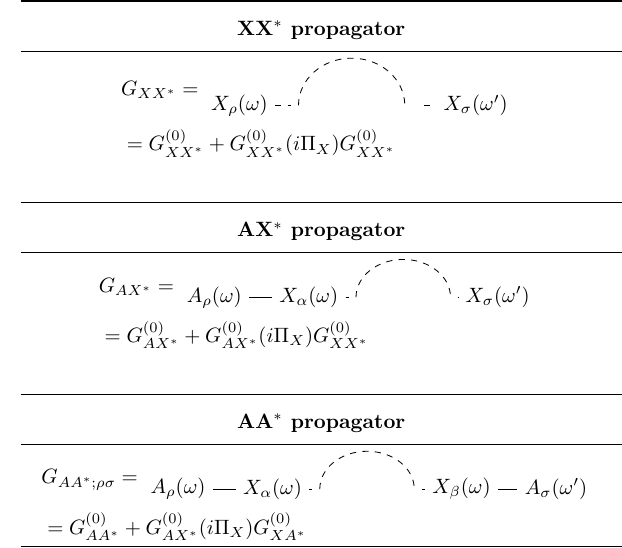}
\caption{The loop corrected propagators of the theory.}
\label{fig:loop_propagators}
\end{figure}

To emphasize how the imprint of the quantum fluctuations dresses each propagator, we 
arrange them in a matrix in the field space as
\begin{equation}
    G(\omega)=
    \begin{pmatrix}
        G_{AA^*}(\omega)&G_{AX^*}(\omega)\\
        G_{XA^*}(\omega)&G_{XX^*}(\omega)
    \end{pmatrix}.
\end{equation}
As the theory mixes fields, the off-diagonal components do not vanish. Each entry 
represents the full propagator, including elements of all orders in the perturbative 
expansion, not only the tree level terms. Next, we invoke the Dyson series to the matrix
\begin{equation}
    G=G^{(0)}+G^{(0)}(i\Pi)G^{(0)}+...,
\end{equation}
which, in index notation, reads
\begin{equation}
\label{eq:dyson}
    G_{ij}=G_{ij}^{(0)}+\sum_{k,l}G_{ik}^{(0)}(i\Pi)_{kl}G_{lj}+...,
    \;\;\;i,j,k,l\in\{\mathbf{A},\mathbf{X}\}.
\end{equation}
However, because there is only one self-correction term, the matrix $\Pi$ looks as
\begin{equation}
    \Pi(\omega)=
    \begin{pmatrix}
        0&0\\
        0&i\Pi_{X}(\omega)
    \end{pmatrix}.
\end{equation}
Therefore, $(i\Pi)_{kl}=\delta_{xk}\delta_{lX}i\Pi_{x}$, and Eq.~(\ref{eq:dyson}) simplifies to
\begin{equation}
    G_{ij}=G_{ij}^{(0)}+G_{iX}^{(0)}(i\Pi_{x})G^{(0)}_{Xj}.
\end{equation}
With this, we are in a position to analyze how the propagators get dressed. The results are captured in the figure \ref{fig:loop_propagators}.\\
It is worth emphasizing that our analysis of loop corrections induced by the interactive nature 
of the internal degrees of freedom has been performed at the level of the bare self-energy tensor. 
As in the familiar spirit of quantum field theories, the loop integral typically diverges. 
Achieving finite, physical observables thus requires renormalization which we leave for 
future exploration. 

\section{Displacement field}
\label{sec:displacement}

In macroscopic quantum electrodynamics of dispersive and absorptive media, 
the displacement field $\mathbf{D}(\omega,\bm{x})$ encodes the medium’s macroscopic 
response to the electromagnetic field. Hence, $\mathbf{D}(\omega, \bm x)$ is the central
quantity that determines how dispersive and dissipative properties of the medium influence
its dielectric response. In anticipation of what we would expect to obtain, let us first recall
how the displacement field is written in classical electromagnetism as a power series in the
applied electric field.

We write the constitutive relation between displacement and electric field in classical optics 
as
\begin{equation}
\mathbf{D}
=\varepsilon_0\mathbf{E}+\mathbf{P}
=\varepsilon_0\!\left[
\varepsilon\,\mathbf{E}
+\chi^{(2)}\!:\mathbf{E}\mathbf{E}
+\chi^{(3)}\!\vdots\mathbf{E}\mathbf{E}\mathbf{E}
+\cdots
\right],
\end{equation}
where $\varepsilon=1+\chi^{(1)}$ is the (generally complex) linear permittivity tensor, and
\begin{equation}
\mathbf{P}
=\varepsilon_0\!\left(
\chi^{(1)}\cdot\mathbf{E}
+\chi^{(2)}\!:\mathbf{E}\mathbf{E}
+\chi^{(3)}\!\vdots\mathbf{E}\mathbf{E}\mathbf{E}
+\cdots
\right).
\end{equation}
Here, we used the colon notation for tensor contractions,
\[
[\chi^{(2)}\!:\mathbf{E}\mathbf{E}]_i
=\chi^{(2)}_{ijk}E_jE_k,\quad
[\chi^{(3)}\!\vdots\mathbf{E}\mathbf{E}\mathbf{E}]_i
=\chi^{(3)}_{ijkl}E_jE_kE_l.
\]
However, when quantizing the medium, integrating out the environmental degrees of freedom
yields an effective action containing a retarded, temporally nonlocal kernel together with a 
corresponding fluctuation term. These ensure consistency with the canonical commutation 
relations and the fluctuation--dissipation theorem. For example, in macroscopic quantum
electrodynamics of linearly responding media, it has been found that the polarization field
gains a noise contribution associated with absorption in the material 
\cite{huttner_quantization_1992,Scheel2008Acta},
\begin{equation}
\mathbf{D} = \varepsilon_0 \varepsilon \mathbf{E} +\mathbf{P}_\text{N} .
\end{equation}
The current theory will provide a generalization to this expression to nonlinearly responding
media where it is not immediately obvious how the quantum noise manifests itself.

When using a Lagrangian formulation of the electromagnetic field, the displacement field is 
formally introduced as
\begin{equation}
  \mathbf{D}(\omega,\bm x)=\frac{\partial \mathcal{L}^{\text{eff}}(\mathbf{E},\mathbf{E^*})}{\partial \mathbf{E^{*}}(\omega,\bm x)}.
\end{equation}
The displacement field can thus be identified as the conjugate variable to the electric field.
Starting from the generating functional $W[\mathbf{J}]$ and defining the effective action by its 
Legendre transform, one obtains $-\delta S^\text{eff} / \delta \mathbf{E}^* =\mathbf{J}$ which 
simplifies to 
\begin{equation}
    \mathbf{D}(\omega,\bm x)=\frac{\delta S^{\text{eff}}}{\delta \mathbf{E^{*}}}
    = -\frac{i\hbar}{Z^{\text{eff}}}
      \frac{\delta Z^{\text{eff}}}{\delta \mathbf{E^{*}}},
      \label{Displacement-effective}
\end{equation}
as shown in detail in Appendix~\ref{Displacement}. To make the above relations explicit, we 
now insert the nonlinear effective action, Eq.~\eqref{eq:Z_with_derivatives}, into 
Eq.~(\ref{Displacement-effective}) in order to obtain the explicit form of the displacement 
field as
\begin{widetext}
\begin{equation}
\begin{aligned}
D_\gamma(\omega,\bm x)
= \varepsilon_0
E_\gamma(\omega,\bm x)&
+\,g(\bm x)\,\Gamma_{\mu\gamma}(\omega,\bm x)\,E_\mu(\omega,\bm x)
\\
+\frac{1}{16Z_{\lambda}}&\Bigg\{\int 
\frac{d\omega_1}{2\pi}\frac{d\omega_3}{2\pi} \frac{d\omega_4}{2\pi}\,
\delta(\omega_1-\omega+\omega_3-\omega_4)\Lambda_{\alpha\gamma\nu\mu}(\bm x)\;
E_\alpha(\omega_1,\bm x)\,
E_\nu(\omega_3,\bm x)\,
E^*_\mu(\omega_4,\bm x)
\\&
+\int \frac{d\omega_1}{2\pi}\frac{d\omega_3}{2\pi} \frac{d\omega_2}{2\pi}\,
\delta(\omega_1-\omega_2+\omega_3-\omega)\Lambda_{\alpha\beta\nu\gamma}(\bm x)\;
E_\alpha(\omega_1,\bm x)\,
E^*_\beta(\omega_2,\bm x)\,
E_\nu(\omega_3,\bm x)
\Bigg\}.
\label{displacement_field}
\end{aligned}
\end{equation}
\end{widetext}
The first term, $\varepsilon_0E_\gamma(\omega,\bm x)$, represents the vacuum 
contribution to the displacement field
in the absence of a medium where $g=0$ and $\Lambda=0$.

The second term, 
$g(\bm x)\,\Gamma_{\mu\gamma}(\omega,\bm x)\,E_\mu(\omega,\bm x)$, adds the 
first-order correction to the displacement field due to medium's linear response, characterized 
by the kernel $\Gamma_{\mu\gamma}$, which consists of both dispersion and absorption.
The linear susceptibility is then defined as the first functional derivative of the displacement 
field with respect to the electric field, i.e.,
\begin{equation}
    \begin{aligned}
    \chi^{(1)}_{\alpha \beta} \coloneqq \chi^{(1,0)}_{\alpha \beta}(\omega;\omega_1)
    &=\frac{1}{\epsilon_0} \frac{\delta D_\alpha(\omega,\bm{x})}{\delta 
   E_\beta(\omega_1,\bm{x}_1)}\bigg |_{{E}=0}-\delta_{\alpha \beta}\\ 
    &= g(\bm x){\varepsilon_0}^{-1}  \Gamma_{\alpha \beta}.       
    \end{aligned}
\end{equation}

One observes that the terms associated with the second-order susceptibility 
($\chi^{(2)}$) vanishes in Eq.~\eqref{displacement_field}. This reflects the fact that
we assumed an isotropic material in which ($\chi^{(2)}$) vanishes due to its spatial inversion
symmetry. The remaining two integrals in the curly brackets in Eq.~\eqref{displacement_field} 
represent the third-order nonlinear polarization contribution arising from the quartic 
self-interaction. These terms are of third order in the electric field and describe four-wave 
mixing processes. 
The delta functions enforce energy conservation among the four relevant frequencies.
The third-order susceptibility can thus be read off as
\begin{equation}
\begin{aligned}
   \chi^{(3)}_{\alpha \beta \mu \nu} &\coloneqq \chi^{(2,1)}_{\alpha \beta \mu \nu}(\omega;\omega_1,\omega_2,\omega_3)\\
   &=\frac{1}{2\epsilon_0}\frac{\delta^3 D_\alpha(\omega,\bm{x})}{\delta 
   E_\beta(\omega_1,\bm{x}_1) \delta E_\mu^*(\omega_2,\bm{x}_2) \delta 
   E_\nu(\omega_3,\bm{x}_3)}\Bigg|_{\raisebox{1.2ex}{$\scriptsize
\substack{
   f = f^{*} = 0 \\
   E = E^{*} = 0
}$}}\\ 
   &=\frac{1}{32\varepsilon_0} (\Lambda_{\alpha \beta \mu \nu} +\Lambda_{\alpha \nu \mu \beta} ),\\
   &= \frac{1}{32}\varepsilon_0^3 \alpha^4 \bigg [ \frac{\lambda_{\gamma \sigma \rho \kappa}}{4!}\Big(\chi^{(1)}_{\alpha \gamma}(\omega_1)\chi^{(1)}_{\beta \sigma}(\omega_2)\chi^{(1)}_{\mu \rho}(\omega_3)\chi^{(1)}_{\nu \kappa}(\omega)\Big)
   \\&\quad+\frac{\lambda_{\gamma \kappa \rho \sigma}}{4!}\Big(\chi^{(1)}_{\alpha \gamma}(\omega_1)\chi^{(1)}_{\nu \kappa}(\omega_2)\chi^{(1)}_{\mu \rho}(\omega_3)\chi^{(1)}_{\beta \sigma}(\omega)\Big)\bigg ]
\end{aligned}
\end{equation}
where the rank-four tensor $\Lambda_{\alpha \beta \mu \nu}$ has been defined in Eq.~(\ref{eq:Lambda}).
As this quantity is directly related to the strength $\lambda_{\alpha \beta \mu \nu}$ of the 
self-interaction, i.e., the anharmonic contribution to the matter field action, 
Eq.~(\ref{S_lambda_freq}), this third-order nonlinearity is solely a consequence of the 
anharmonicity of the matter fields. Moreover, the tensor $\Lambda_{\alpha \beta \mu \nu}$ contains a 
product of four factors of the linear susceptibility $\chi^{(1)}_{\alpha \beta }$ as expected from
Miller's rule \cite{boyd_nonlinear_2020,Miller1964}.
This result constitutes the first step in the quantization program of electromagnetic fields
in nonlinearly responding dielectric matter with dispersion and absorption.

\section{Conclusions}
\label{sec:conclusions}

Starting from the mesoscopic harmonic oscillator model of the matter polarization coupled
to an harmonic oscillator reservoir, we have extended the model to include a quartic
self-interaction of the matter fields, that can be interpreted as an anharmonic contribution
of the matter to an externally applied electric field. We used the path-integral formalism
to obtain the effective action resulting from integrating out the reservoir degrees of freedom.
The integration over the matter fields is done in two steps: first, by integrating out the 
bilinear contributions resulting in a Gaussian functional integral, and then applying
perturbation theory by replacing the source fields by functional derivatives. The resulting
generating function $Z_\text{NL}[\mathbf{f},\mathbf{J}]$ then contains perturbative
nonlinear contributions resulting from the matter-field self-interaction. With the generating function at hand, we constructed the relevant Feynman diagrams both to tree level and one-loop
corrections. 

As an immediate consequence, we find the third-order nonlinear susceptibility $\chi^{(3)}$
as a function of the coupling constant $\lambda_{ijkl}$ in the matter self-interaction and
a product of linear response kernels $\Gamma_{ij}$. This way, it is possible to construct
dispersive and absorbing nonlinear susceptibilities from lower-order susceptibilities. In 
particular, in isotropic media, this involves only a small number of fitting parameters.

With the generating function at hand, one can start constructing expectation values of operator 
products. In particular, from the one-loop corrected vertex and propagators, the photon-photon 
propagator (Green's function) can be seen as resulting from a vector potential that has been
corrected by a $\chi^{(3)}$ contribution. This will in effect enable one to construct a 
theory of nonlinear macroscopic quantum electrodynamics, that includes a nonlinear noise
polarization such that the nonlinear FDT is satisfied, which we leave for future work. 

\acknowledgments
This work was funded by the Deutsche Forschungsgemeinschaft (DFG, German Research 
Foundation) through IRTG 2676/1 ‘Imaging of Quantum Systems’, project no. 437567992. M.O. 
acknowledges the financial support from the Photonics Research and Innovation Flagship 
(PREIN - decision 320165) and the Research Council of Finland project AQUA-PHOT (decision 
Grant No. 349350).

\appendix
\begin{widetext}
\section{Functional derivatives of the generatingpartition function}
\label{sec:Functional_derivatives}
We first compute the Gaussian contribution $Z_{\text{Gauss}}$ explicitly in the time 
domain. For this purpose, we adopt the following Fourier convention
\begin{align}
     \nonumber
     \mathbf{B}(t,\bm x) &= \int\frac{d\omega}{2\pi}\, e^{i\omega t}\, \mathbf{B}(\omega,\bm x),\\[4pt]
     (\Gamma * \mathbf{B})(t) &= \int\frac{d\omega}{2\pi}\, e^{i\omega t}\, \Gamma(\omega,\bm x)\, \mathbf{B}(\omega,\bm x),
\end{align}
where $\mathbf{B} = (\mathbf{f},\, \dot{\mathbf{A}})$. 
Substituting these expressions into the definition of $Z_{\text{Gauss}}$, we obtain
\begin{equation}
\begin{aligned}
Z_{\text{Gauss}}[\mathbf{E},\mathbf{f}]
= \exp\Bigg[
  \int \! dt\, d^{3}x\,
  \frac{d\omega}{2\pi}\,
  \frac{d\omega'}{2\pi}\,
  e^{-i(\omega+\omega')t}
  \Bigl[
  E_{\mu}(\omega)\,\Gamma_{\mu\nu}(\omega')\,E_{\nu}(\omega')
  + {f}_{\mu}(\omega)\,\Gamma_{\mu\nu}(\omega')\,{f}_{\nu}(\omega')
\\
  - {f}_{\mu}(\omega)\,\Gamma_{\mu\nu}(\omega')\,E_{\nu}(\omega')
  - E_{\mu}(\omega)\,\Gamma_{\mu\nu}(\omega')\,{f}_{\nu}(\omega')
  \Bigr]
\Bigg].
\end{aligned}
\end{equation}
Because the time integral commutes with the frequency integrals, 
the exponential factor $\int dte^{-i(\omega+\omega')t}$ produces a Dirac delta $\delta(\omega+\omega')$. 
Using the reality condition $\mathbf{B}(-\omega) = \mathbf{B}^{*}(\omega)$, the expression simplifies to
\begin{equation}
\begin{aligned}
Z_{\text{Gauss}}[\mathbf{E},\mathbf{f}]
=&
\exp\Bigg[
\frac{i}{2\hbar}
\int \frac{d\omega_{1}}{2\pi}\, d^{3}x_{1}\,
g(\mathbf{x}_{1}) \, 
\big( f_{\mu}(\omega_{1},\mathbf{x}_{1})
      - E_{\mu}(\omega_{1},\mathbf{x}_{1}) \big)
 \, \Gamma_{\mu\nu}(\omega_{1},\mathbf{x}_{1})
 \,
\big( f_{\nu}^{*}(\omega_{1},\mathbf{x}_{1})
      - E_{\nu}^{*}(\omega_{1},\mathbf{x}_{1}) \big)
\Bigg] .
\end{aligned}
\end{equation}
For the self-interaction term, we can proceed in complete analogy with the previous case. \\
To calculate the functional derivatives in Eq.~\eqref{eq:Z_with_derivatives}, it is perhaps 
more useful to drop the $Z_{\text{e.m.}}$ and $Z_{\text{s}}$ and only concentrate on the 
sector that depends on the matter field,

 \begin{equation}
    \begin{aligned}
        \frac{\delta Z_{\text{Gauss}}}{\delta f_{\gamma}(\omega_{2},\bm{x}_{2})}
        &= Z_{\text{Gauss}}\left(\frac{i}{2\hbar}\right)
        \int \frac{d\omega_{1}}{2\pi}\, d^{3}x_{1}\,
        \Big(\delta_{\gamma\mu}\,\delta^{(1)}(\omega_{1}-\omega_{2})\,\delta^{(3)}(\bm{x}_{1}-\bm{x}_{2})\,
        \Gamma_{\mu\nu}(\omega_{1},\bm{x}_{1})\, f_{\nu}^{*}(\omega_{1},\bm{x}_{1})\\&\qquad-\,\delta_{\gamma\mu}\,\delta^{(1)}(\omega_{1}-\omega_{2})\,\delta^{(3)}(\bm{x}_{1}-\bm{x}_{2})\,\Gamma_{\mu\nu}(\omega_{1},\bm{x}_{1})\, E_{\nu}^{*}(\omega_{1},\bm{x}_{1})\Big)\\
        &=Z_{\text{Gauss}}\left(\frac{i}{2\hbar}\right){g}\Gamma_{\gamma\nu}(\omega_{2},\bm{x}_{2})\left(f_{\nu}^{*}(\omega_{2},\bm{x}_{2})-{\alpha}
        E_{\nu}^{*}(\omega_{2},\bm{x}_{2})\right)\equiv Z_{\text{Gauss}} 
        \oscirc{\gamma}^{(2)}
    \end{aligned}
\end{equation}   
This result is a vector whose component is the free index $\gamma$, and the superscript refers to the label of the parameters $\omega_2$ and $\mathbf{x}_2$. As it appears quite often 
throughout the calculation, we abbreviate it using the name $\oscirc{\gamma}$. It is also 
worth noting that
\begin{equation}
        \frac{\delta Z_{\text{Gauss}}}{\delta f_{\gamma}^{*}(\omega_{2},\bm{x}_{2})}= 
        Z_{\text{Gauss}}{g}\left(\frac{i}{2\hbar}\right)\left(f_{\nu}(\omega_{2},
        \bm{x}_{2})-{\alpha}E_{\nu}(\omega_{2},\bm{x}_{2})\right)
        \Gamma_{\gamma\nu}(\omega_{2},\bm{x}_{2})=Z_{\text{Gauss}}\oscirc{\gamma}
        ^{\hspace{- 0.5em}{*}}\equiv Z_{\text{Gauss}}\osstar{\gamma}^{(2)}.
\end{equation}
The higher-order derivatives give the following results:
    \begin{equation}
        \begin{aligned}
              \frac{\delta^{2}Z_{\text{Gauss}}}{\delta f_{\gamma}\delta f^{*}_{\sigma}(\omega_3,\mathbf{x}_3)}=Z_{\text{Gauss}}\osstar{\sigma}^{(3)}\oscirc{\gamma}^{(2)}+Z_{\text{Gauss}}\left(\frac{i}{2\hbar}\right)\Gamma_{\gamma\sigma}(\omega_2,\mathbf{x}_2)\delta^{(1)}(\omega_{2}-\omega_3)\delta^{(3)}(\bm{x}_2-\bm{x}_3)g(\mathbf{x}_2),
        \end{aligned}
    \end{equation}
\begin{equation}
\begin{aligned}
\frac{\delta^{3} Z_{\mathrm{Gauss}}}
     {\delta f_{\gamma}\,\delta f^{*}_{\sigma}\,\delta f_{\lambda}(\omega_{4},\bm{x}_{4})}
= Z_{\mathrm{Gauss}} \Bigl[
  \oscirc{\lambda}^{(4)} \osstar{\sigma}^{(3)} \oscirc{\gamma}^{(2)}
  {}+ \oscirc{\lambda}^{(4)} \left(\frac{i}{2\hbar}\right) \Gamma_{\gamma\sigma}(\omega_2,\bm{x}_2)
     \delta^{(1)}(\omega_{2}-\omega_{3}) \delta^{(3)}(\bm{x}_{2}-\bm{x}_{3}) g(\bm{x}_{2})
\\
  {}+ g(\bm{x}_{3}) \left(\frac{i}{2\hbar}\right)
     \delta^{(1)}(\omega_{3}-\omega_{4}) \delta^{(3)}(\bm{x}_{3}-\bm{x}_{4})
     \Gamma_{\lambda\sigma}(\omega_3,\bm{x}_3)\, \oscirc{\gamma}^{(2)}
\Bigr],
\end{aligned}
\end{equation}

\begin{equation}
\begin{aligned}
\frac{\delta^{4}Z_{\text{Gauss}}}{\delta f_{\gamma}\delta f^{*}_{\rho}\delta f_{\lambda}\delta 
f^{*}_{\tau}(\omega_{5},\bm{x}_{5})}=Z_{\text{Gauss}}\big[\osstar{\tau}^{(5)}\, 
\oscirc{\lambda}^{(4)}\,\osstar{\rho}^{(3)}\,\oscirc{\gamma}^{(2)}
+\left(\frac{i}{2\hbar}\right)\osstar{\tau}^{(5)}\,\oscirc{\lambda}^{(4)}\,
\Gamma_{\gamma\sigma}(\omega_2,\bm{x}_2)\delta^{(1)}(\omega_{2}-\omega_{3})\delta^{(3)}(\bm{x}_2-\bm{x}_{3})+\\
+\left(\frac{i}{2\hbar}\right)\osstar{\tau}^{(5)}
\oscirc{\gamma}^{(2)}\Gamma_{\lambda\sigma}(\omega_3,\bm{x}_3)\delta^{(1)}(\omega_{3}-\omega_{4})\delta^{(3)}
(\bm{x}_{3}-\bm{x}_4)
+\left(\frac{i}{2\hbar}\right)\osstar{\sigma}^{(3)}
\oscirc{\gamma}^{(2)}\Gamma_{\lambda\tau}(\omega_4,\bm{x}_4)\delta^{(1)}(\omega_{4}-\omega_{5})
\delta^{(3)}(\bm{x}_{4}-\bm{x}_{5}) +\\
+\left(\frac{i}{2\hbar}\right)\oscirc{\lambda}^{(4)}
\osstar{\sigma}^{(3)}\Gamma_{\gamma\tau}(\omega_2,\bm{x}_2)\delta^{(1)}(\omega_2-\omega_5)
\delta^{(3)}(\bm{x}_2-\bm{x}_{5}) +\\
+\left(\frac{i}{2\hbar}\right)^{2}\Gamma_{\lambda\tau}(\omega_2,\bm{x}_2)\Gamma_{\gamma\sigma}(\omega_4,\bm{x}_4)\delta^{(1)}
(\omega_4-\omega_5)\delta^{(3)}(\bm{x}_4-\bm{x}_5)\delta^{(1)}(\omega_2-\omega_3)\delta^{(3)}
(\bm{x}_2-\bm{x}_{3}) +\\
+\left(\frac{i}{2\hbar}\right)^{2}\Gamma_{\lambda\sigma}(\omega_2,\bm{x}_2)\Gamma_{\gamma\tau}(\omega_3,\bm{x}_3)\delta^{(1)}
(\omega_3-\omega_4)\delta^{(3)}(\bm{x}_3-\bm{x}_4)\delta^{(1)}(\omega_2-\omega_5)\delta^{(3)}
(\bm{x}_2-\bm{x}_5)\big].
\end{aligned}
\end{equation}

Finally, note that the constraints imposed by the Dirac delta functions on each term ensure 
that all arguments are evaluated at the same frequency. The chain of delta functions enforces 
frequency equality, making these terms proportional to $\delta(0)$. They will contain no legs 
on the Feynman diagram representation, since they do not allow external energy flow. These are 
the so-called vacuum bubbles. They belong to the class of disconnected diagrams. We can safely 
drop them by noting that the physical information of the theory is revealed through the
generator of the connected diagrams. By expanding the remaining product tensor we recover the 
expression in Eq.~\eqref{eq:Zeff}.
\end{widetext}
\section{Displacement field in the Lagrangian formulation}
\label{Displacement}
For this profit is better to work directly with the electric field instead. When using a 
Lagrangian formulation of the electromagnetic field, the displacement field is formally 
introduced as
\begin{equation}
    \mathbf{D}=\frac{\partial\mathcal{L}_{eff}[\mathbf{E,\mathbf{E}^{*}}]}{\partial\mathbf{E}^{*}}
\end{equation}
This definition can be extended to a path-integral framework.

Motivated by the aim of removing the dependence on the unphysical source variable and expressing 
the theory in terms of the physical quantities, i.e. the fields, we introduce the effective 
action. It can be defined through a Legendre transformation as
\begin{equation}    
 S_{\text{eff}}[\mathbf{E},\mathbf{E}^*]=W[\mathbf{J},\mathbf{J}^*]-\int dtd^{3}x
 (\mathbf{J}^{*}\cdot\mathbf{E}-\mathbf{J}\cdot\mathbf{E}^{*}).
\end{equation}
With that equation in hand, we can differentiate it to get
\begin{equation}
    \delta S_{\text{eff}}=\delta W-\int\frac{d\omega}{2\pi}\left[-\delta \mathbf{J}\cdot\mathbf{E}^{*}- \mathbf{J}\cdot\delta\mathbf{E}^{*}+\delta \mathbf{J}^{*}\cdot\mathbf{E}+ \mathbf{J}^{*}\cdot\delta\mathbf{E}\right]
\end{equation}
and, using the relation $\frac{\delta W}{\delta \mathbf{J}}=\mathbf{E}^{*}$ and $\frac{\delta W}{\delta \mathbf{J}^{*}}=\mathbf{E}$, we recover the standard correspondence between the effective action $\Gamma[\mathbf{E}]$ and the generating functional $W[\mathbf{J}]$:
\begin{equation}
    \frac{ \delta S_{\text{eff}}}{\delta\mathbf{E}^{*}}=-\mathbf{J}\;\;\;\;\;\frac{\delta S_{\text{eff}}}{\delta\mathbf{E}}=+\mathbf{J}^{*}.
    \label{eq:Source_Legendre}
\end{equation}
The connected generating functional $W$ is customarily defined through the logarithm of the generating function, and with this equivalence, the following relation naturally emerges.
\begin{equation}
    \frac{\delta W}{\delta \mathbf{J}} = -\frac{i\hbar}{Z_{\text{eff}}}\frac{\delta Z_{\text{eff}}}{\delta \mathbf{J}}
\end{equation}
Similarly, by considering the definition of the generating function as (\ref{eq:definition_Z_eff}), an analog expression for the electromagnetic field emerges:
\begin{equation}
    -\frac{i\hbar}{Z_{\text{eff}}}\frac{\delta Z_{\text{eff}}}{\delta \mathbf{E}^{*}}=\frac{\delta S_{\text{eff}}}{\delta\mathbf{E}^{*}}.
\end{equation}

This result holds specifically due to the way the effective generating functional is defined,
namely by integrating out the fields as a direct consequence of the Legendre transformation.

To finally see the relation with the Lagrangian, let us vary the effective action with respect 
to $\mathbf{E}^{*}$:
\begin{equation}
    \delta S_{\text{eff}}=\int dtd^{3}x\left[\frac{\partial\mathcal{L}_{eff}}{\partial\mathbf{E}^{*}}\delta\mathbf{E}^{*}+\frac{\mathcal{L}_{eff}}{\partial(\partial_{\mu}\mathbf{E}^{*})}\delta(\partial_{\mu}\mathbf{E}^{*})+\mathcal{O}(\partial^{2})\right]
\end{equation}
After integrating by parts the second term and by using the definition of the functional 
derivative, it can be verified that
\begin{equation}
     \frac{\delta S_{\text{eff}}}{\delta \mathbf{E}}
    = \frac{\partial \mathcal{L}_{\text{eff}}}{\partial \mathbf{E}}
      - \partial_{\mu}  \left(
          \frac{\partial \mathcal{L}_{\text{eff}}}{\partial (\partial_{\mu} \mathbf{E})}
        \right).
\end{equation}
In cases where the Lagrangian does not depend on the derivative of the field, or when such a dependence reduces to a constant, the second term vanishes. Hence,
\begin{equation}
    -\frac{i\hbar}{Z_{\text{eff}}}\frac{\delta Z_{\text{eff}}}{\delta \mathbf{J}}
    = \frac{\delta S_{\text{eff}}}{\delta \mathbf{E}^{*}}
    = \frac{\partial \mathcal{L}_{\text{eff}}[\mathbf{E}, \mathbf{E}^{*}]}{\partial \mathbf{E}^{*}}
    = \mathbf{D}.
    \label{eq:Displacement}
\end{equation}
This establishes the formal correspondence between the displacement field and the derivative of 
the effective Lagrangian with respect to the complex-conjugate electric field.

It is also worth noting that, because of Eqs.~(\ref{eq:Source_Legendre}) 
and~(\ref{eq:Displacement}), the displacement field mathematically acts as a source. Formally, 
it is the source conjugate to $\mathbf{E}^{*}$; that is, it plays the role of the effective 
source that would reproduce the corresponding field configuration. In this way, the equation of 
motion of the effective action coincides with the macroscopic Maxwell equation in the presence 
of an external source.
 
\bibliographystyle{apsrev4-2} 
\bibliography{references.bib}

@article{huttner_quantization_1992,
	title = {Quantization of the electromagnetic field in dielectrics},
	volume = {46},
	rights = {http://link.aps.org/licenses/aps-default-license},
	issn = {1050-2947, 1094-1622},
	url = {https://link.aps.org/doi/10.1103/PhysRevA.46.4306},
	doi = {10.1103/PhysRevA.46.4306},
	pages = {4306--4322},
	number = {7},
	journal = {Physical Review A},
	shortjournal = {Phys. Rev. A},
	author = {Huttner, Bruno and Barnett, Stephen M.},
	urldate = {2024-11-08},
	date = {1992-10-01},
    month = oct,
	year = {1992},
	langid = {english},
}

@article{bechler_quantum_1999,
	title = {Quantum electrodynamics of the dispersive dielectric medium–a path integral approach},
	volume = {46},
	issn = {0950-0340, 1362-3044},
	url = {http://www.tandfonline.com/doi/abs/10.1080/09500349908231312},
	doi = {10.1080/09500349908231312},
	language = {en},
	number = {5},
	urldate = {2025-03-13},
	journal = {Journal of Modern Optics},
	author = {Bechler, Adam},
	month = apr,
	year = {1999},
	pages = {901--921},
}

@article{HilleryMlodinow1997,
  title = {Quantized fields in a nonlinear dielectric medium: A microscopic approach},
  author = {Hillery, Mark and Mlodinow, Leonard},
  journal = {Phys. Rev. A},
  volume = {55},
  issue = {1},
  pages = {678--689},
  numpages = {0},
  year = {1997},
  month = {Jan},
  publisher = {American Physical Society},
  doi = {10.1103/PhysRevA.55.678},
  url = {https://link.aps.org/doi/10.1103/PhysRevA.55.678}
}

@article{ScheelWelsch2006,
  title = {Quantum Theory of Light and Noise Polarization in Nonlinear Optics},
  author = {Scheel, Stefan and Welsch, Dirk-Gunnar},
  journal = {Phys. Rev. Lett.},
  volume = {96},
  issue = {7},
  pages = {073601},
  numpages = {4},
  year = {2006},
  month = {Feb},
  publisher = {American Physical Society},
  doi = {10.1103/PhysRevLett.96.073601},
  url = {https://link.aps.org/doi/10.1103/PhysRevLett.96.073601}
}

@article{ScheelWelsch2006b,
doi = {10.1088/0953-4075/39/15/S17},
url = {https://doi.org/10.1088/0953-4075/39/15/S17},
year = {2006},
month = {jul},
publisher = {},
volume = {39},
number = {15},
pages = {S711},
author = {Scheel, S and Welsch, D-G},
title = {Causal nonlinear quantum optics},
journal = {Journal of Physics B: Atomic, Molecular and Optical Physics}
}

@article{Lindel2021,
  title = {Macroscopic quantum electrodynamics approach to nonlinear optics and application to polaritonic quantum-vacuum detection},
  author = {Lindel, Frieder and Bennett, Robert and Buhmann, Stefan Yoshi},
  journal = {Phys. Rev. A},
  volume = {103},
  issue = {3},
  pages = {033705},
  numpages = {18},
  year = {2021},
  month = {Mar},
  publisher = {American Physical Society},
  doi = {10.1103/PhysRevA.103.033705},
  url = {https://link.aps.org/doi/10.1103/PhysRevA.103.033705}
}

@book{OrnigottiBook,
author = {Ornigotti, Marco},
title = {A Field Theory Approach to Photonics},
publisher = {IOP Publishing},
year = {2025},
series = {2053-2563},
isbn = {978-0-7503-5789-0},
url = {https://doi.org/10.1088/978-0-7503-5789-0},
doi = {10.1088/978-0-7503-5789-0}
}

@article{Ornigotti2019,
  title = {Path-integral description of quantum nonlinear optics in arbitrary media},
  author = {Difallah, Mosbah and Szameit, Alexander and Ornigotti, Marco},
  journal = {Phys. Rev. A},
  volume = {100},
  issue = {5},
  pages = {053845},
  numpages = {12},
  year = {2019},
  month = {Nov},
  publisher = {American Physical Society},
  doi = {10.1103/PhysRevA.100.053845},
  url = {https://link.aps.org/doi/10.1103/PhysRevA.100.053845}
}

@article{Scheel1998,
  title = {QED commutation relations for inhomogeneous Kramers-Kronig dielectrics},
  author = {Scheel, Stefan and Kn\"oll, Ludwig and Welsch, Dirk-Gunnar},
  journal = {Phys. Rev. A},
  volume = {58},
  issue = {1},
  pages = {700--706},
  numpages = {0},
  year = {1998},
  month = {Jul},
  publisher = {American Physical Society},
  doi = {10.1103/PhysRevA.58.700},
  url = {https://link.aps.org/doi/10.1103/PhysRevA.58.700}
}

@article{Dung1998,
  title = {Three-dimensional quantization of the electromagnetic field in dispersive and absorbing inhomogeneous dielectrics},
  author = {Dung, Ho Trung and Kn\"oll, Ludwig and Welsch, Dirk-Gunnar},
  journal = {Phys. Rev. A},
  volume = {57},
  issue = {5},
  pages = {3931--3942},
  numpages = {0},
  year = {1998},
  month = {May},
  publisher = {American Physical Society},
  doi = {10.1103/PhysRevA.57.3931},
  url = {https://link.aps.org/doi/10.1103/PhysRevA.57.3931}
}

@article{Gruner1996,
  title = {Green-function approach to the radiation-field quantization for homogeneous and inhomogeneous Kramers-Kronig dielectrics},
  author = {Gruner, T. and Welsch, D.-G.},
  journal = {Phys. Rev. A},
  volume = {53},
  issue = {3},
  pages = {1818--1829},
  numpages = {0},
  year = {1996},
  month = {Mar},
  publisher = {American Physical Society},
  doi = {10.1103/PhysRevA.53.1818},
  url = {https://link.aps.org/doi/10.1103/PhysRevA.53.1818}
}

@article{Scheel2008Acta,
  title={Macroscopic quantum electrodynamics-concepts and applications},
  author={Scheel, Stefan and Buhmann, Stefan Yoshi and others},
  journal={Acta Phys. Slovaca},
  volume={58},
  number={5},
  pages={675--809},
  year={2008}
}

@book{Buhmann2013I,
  title={Dispersion Forces I: Macroscopic quantum electrodynamics and ground-state Casimir, Casimir--Polder and van der Waals forces},
  author={Buhmann, Stefan Yoshi},
  volume={247},
  year={2013},
  publisher={Springer}
}

@article{Suttorp2004,
  title = {Field quantization in inhomogeneous absorptive dielectrics},
  author = {Suttorp, L. G. and Wubs, Martijn},
  journal = {Phys. Rev. A},
  volume = {70},
  issue = {1},
  pages = {013816},
  numpages = {18},
  year = {2004},
  month = {Jul},
  publisher = {American Physical Society},
  doi = {10.1103/PhysRevA.70.013816},
  url = {https://link.aps.org/doi/10.1103/PhysRevA.70.013816}
}

@article{Tip2001,
  title = {Equivalence of the Langevin and auxiliary-field quantization methods for absorbing dielectrics},
  author = {Tip, A. and Kn\"oll, L. and Scheel, S. and Welsch, D.-G.},
  journal = {Phys. Rev. A},
  volume = {63},
  issue = {4},
  pages = {043806},
  numpages = {7},
  year = {2001},
  month = {Mar},
  publisher = {American Physical Society},
  doi = {10.1103/PhysRevA.63.043806},
  url = {https://link.aps.org/doi/10.1103/PhysRevA.63.043806}
}

@article{Schmidt1998,
author = {Eduard Schmidt and John Jeffers and Stephen M. Barnett and Ludwig Knöll and Dirk-Gunnar Welsch},
title = {Quantum theory of light in nonlinear media with dispersion and absorption},
journal = {Journal of Modern Optics},
volume = {45},
number = {2},
pages = {377--401},
year = {1998},
publisher = {Taylor \& Francis},
doi = {10.1080/09500349808231696},
URL = {https://doi.org/10.1080/09500349808231696},
eprint = {https://doi.org/10.1080/09500349808231696}
}

@article{Matloob1995,
  title = {Electromagnetic field quantization in absorbing dielectrics},
  author = {Matloob, Reza and Loudon, Rodney and Barnett, Stephen M. and Jeffers, John},
  journal = {Phys. Rev. A},
  volume = {52},
  issue = {6},
  pages = {4823--4838},
  numpages = {0},
  year = {1995},
  month = {Dec},
  publisher = {American Physical Society},
  doi = {10.1103/PhysRevA.52.4823},
  url = {https://link.aps.org/doi/10.1103/PhysRevA.52.4823}
}

@article{Kaminer_review,
    author = {Rivera, Nicholas and Kaminer, Ido},
    title = {Light–matter interactions with photonic quasiparticles},
    journal = {Nature Reviews Physics},
    year = {2020},
    volume = {2},
    pages = {538--561},
    doi = {10.1038/s42254-020-0224-2},
    url = {https://doi.org/10.1038/s42254-020-0224-2}
}

@article{Heinz2002,
  title = {Generalized fluctuation-dissipation theorem for nonlinear response functions},
  author = {Wang, Enke and Heinz, Ulrich},
  journal = {Phys. Rev. D},
  volume = {66},
  issue = {2},
  pages = {025008},
  numpages = {17},
  year = {2002},
  month = {Jul},
  publisher = {American Physical Society},
  doi = {10.1103/PhysRevD.66.025008},
  url = {https://link.aps.org/doi/10.1103/PhysRevD.66.025008}
}

@article{feynman_vernon_1963,
  author    = {R. P. Feynman and F. L. Vernon},
  title     = {The theory of a general quantum system interacting with a linear dissipative system},
  journal   = {Annals of Physics},
  volume    = {24},
  pages     = {118--173},
  year      = {1963},
  doi       = {10.1016/0003-4916(63)90068-X}
}

@article{toll_1956,
  author    = {J. S. Toll},
  title     = {Causality and the Dispersion Relation: Logical Foundations},
  journal   = {Physical Review},
  volume    = {104},
  number    = {6},
  pages     = {1760--1770},
  year      = {1956},
  doi       = {10.1103/PhysRev.104.1760}
}

@article{hopfield1958theory,
  title        = {Theory of the Contribution of Excitons to the Complex Dielectric Constant of Crystals},
  author       = {Hopfield, J. J.},
  journal      = {Physical Review},
  volume       = {112},
  number       = {5},
  pages        = {1555--1567},
  year         = {1958},
  doi          = {10.1103/PhysRev.112.1555}
}

@article{Weyl1929,
  author       = {Hermann Weyl},
  title        = {Elektron und Gravitation. I},
  journal      = {Zeitschrift f{\"u}r Physik},
  volume       = {56},
  pages        = {330--352},
  year         = {1929},
  doi          = {10.1007/BF01339504},
  url          = {https://link.springer.com/article/10.1007/BF01339504},
  language     = {German},
  note         = {Introduces the modern concept of local gauge invariance (Eichinvarianz).}
}

@article{ORaifeartaigh2000,
  author       = {Lochlainn O’Raifeartaigh and Norbert Straumann},
  title        = {Gauge theory: Historical origins and some modern developments},
  journal      = {Reviews of Modern Physics},
  volume       = {72},
  number       = {1},
  pages        = {1--23},
  year         = {2000},
  doi          = {10.1103/RevModPhys.72.1},
}

@article{Fermi1932,
  author  = {Fermi, Enrico},
  title   = {Quantum Theory of Radiation},
  journal = {Reviews of Modern Physics},
  volume  = {4},
  number  = {1},
  pages   = {87--132},
  year    = {1932},
  doi     = {10.1103/RevModPhys.4.87}
}

@book{Bloembergen1965,
  author    = {Nicolaas Bloembergen},
  title     = {Nonlinear Optics},
  publisher = {W. A. Benjamin, Inc.},
  address   = {New York},
  year      = {1965},
  note      = {First edition},
  url       = {https://books.google.com/books/about/Nonlinear_Optics.html?id=zqcYAQAAMAAJ}
}

@book{ItzyksonZuber1980,
  author    = {C. Itzykson and J.-B. Zuber},
  title     = {Quantum Field Theory},
  publisher = {McGraw-Hill},
  year      = {1980}
}

@article{soltani2008,
  author  = {Kheirandish, F. and Soltani, M.},
  title   = {Extension of the Huttner-Barnett model to a magnetodielectric medium},
  journal = {Physical Review A},
  volume  = {78},
  pages   = {012102},
  year    = {2008},
}

@article{saravi,
	title={Nonperturbative theory of spontaneous parametric down-conversion in open and dispersive optical systems},
	author={Krsti{\'c}, Aleksa and Setzpfandt, Frank and Saravi, Sina},
	journal={Physical Review Research},
	volume={5},
	number={4},
	pages={043228},
	year={2023},
	publisher={APS}
}

@article{ornigotti2025,
  author  = {Dal Negro, L. and Franchi, R. and Ornigotti, M.},
  title   = {Nonlinear quantum electrodynamics of epsilon-near-zero nanostructures},
  journal = {Physical Review B},
  volume  = {112},
  pages   = {165433},
  year    = {2025},
}

@article{ornigotti2024,
  author  = {Tamashevich, Y. and Shubitidze, T. and Dal Negro, L. and Ornigotti, M.},
  title   = {Field theory description of the non-perturbative optical
nonlinearity of epsilon-near-zero media},
  journal = {APL Photonics},
  volume  = {9},
  pages   = {016105},
  year    = {2024},
}

@article{reshef2017,
	title={Beyond the perturbative description of the nonlinear optical response of low-index materials},
	author={Reshef, Orad and Giese, Enno and Alam, M Zahirul and De Leon, Israel and Upham, Jeremy and Boyd, Robert W},
	journal={Optics letters},
	volume={42},
	number={16},
	pages={3225--3228},
	year={2017},
	publisher={Optical Society of America}
}

@article{ref46,
  title={Extremely large nondegenerate nonlinear index and phase shift in epsilon-near-zero materials},
  author={Benis, Sepehr and Munera, Natalia and Faryadras, Sanaz and Van Stryland, Eric W and Hagan, David J},
  journal={Optical Materials Express},
  volume={12},
  number={10},
  pages={3856--3871},
  year={2022},
  publisher={Optica Publishing Group}
}

@article{ref47,
  title={Enhanced nonlinear refractive index in $\varepsilon$-near-zero materials},
  author={Caspani, Lucia and Kaipurath, RPM and Clerici, Matteo and Ferrera, Marcello and Roger, Thomas and Kim, J and Kinsey, Nathaniel and Pietrzyk, Monika and Di Falco, Andrea and Shalaev, Vladimir M and others},
  journal={Physical review letters},
  volume={116},
  number={23},
  pages={233901},
  year={2016},
  publisher={APS}
}

@article{ref48,
  title={Extended Drude model for intraband-transition-induced optical nonlinearity},
  author={Wang, Heng and Du, Kang and Jiang, Chuhao and Yang, Zhiqiang and Ren, Lixia and Zhang, Wending and Chua, Soo Jin and Mei, Ting},
  journal={Physical Review Applied},
  volume={11},
  number={6},
  pages={064062},
  year={2019},
  publisher={APS}
}

@article{scalora1,
author = {M. Scalora and J. Trull and C. Cojocaru and M. A. Vincenti and L. Carletti and D. de Ceglia and N. Akozbek and C. De Angelis},
journal = {J. Opt. Soc. Am. B},
number = {8},
pages = {2346--2351},
publisher = {Optica Publishing Group},
title = {Resonant, broadband, and highly efficient optical frequency conversion in semiconductor nanowire gratings at visible and UV wavelengths},
volume = {36},
month = {Aug},
year = {2019},
url = {https://opg.optica.org/josab/abstract.cfm?URI=josab-36-8-2346},
doi = {10.1364/JOSAB.36.002346},
}

@article{scalora2,
author = {M. Scalora and M. A. Vincenti and D. de Ceglia and C. M. Cojocaru and M. Grande and J. W. Haus},
journal = {J. Opt. Soc. Am. B},
number = {10},
pages = {2129--2138},
publisher = {Optica Publishing Group},
title = {Nonlinear Duffing oscillator model for third harmonic generation},
volume = {32},
month = {Oct},
year = {2015},
url = {https://opg.optica.org/josab/abstract.cfm?URI=josab-32-10-2129},
doi = {10.1364/JOSAB.32.002129},
}

@article{scalora3,
    author = {Rodríguez-Suné, L. and Scalora, M. and Johnson, A. S. and Cojocaru, C. and Akozbek, N. and Coppens, Z. J. and Perez-Salinas, D. and Wall, S. and Trull, J.},
    title = {Study of second and third harmonic generation from an indium tin oxide nanolayer: Influence of nonlocal effects and hot electrons},
    journal = {APL Photonics},
    volume = {5},
    number = {1},
    pages = {010801},
    year = {2020},
    month = {01},
    issn = {2378-0967},
    doi = {10.1063/1.5129627},
    url = {https://doi.org/10.1063/1.5129627}
}

@book{boyd_nonlinear_2020,
	address = {London},
	edition = {Fourth edition},
	title = {Nonlinear optics},
	isbn = {978-0-12-811002-7 978-0-12-811003-4},
	language = {english},
	publisher = {Elsevier, AP Academic Press},
	author = {Boyd, Robert W.},
	year = {2020},
	doi = {10.1016/C2015-0-05510-1}
}

@article{Miller1964,
    author = {R. C. Miller},
    title = {Optical second harmonic generation in piezoelectric crystals},
    journal = {Appl. Phys. Lett.},
    year = {1964},
    volume = {5},
    pages = {17-19},
    url = {https://doi.org/10.1063/1.1754022},
    doi = {10.1063/1.1754022}
}

\end{document}